\documentclass[twocolumn]{aastex631}
\usepackage{graphicx}	
\usepackage[fleqn]{amsmath}
\usepackage{eqnarray}
\usepackage{amssymb}	
\usepackage{bm}		    
\usepackage[T1]{fontenc}
\usepackage{natbib}
\usepackage{listings}
\usepackage{multirow}
\usepackage{diagbox}

\received{\today}
\revised{-}
\accepted{-}

\submitjournal{ApJL}

\shorttitle{Fine-Tuned Supernova or Failed Explosion?}

\shortauthors{Regály, Fröhlich \& Vinkó}

\graphicspath{{./}{figures/}}

\begin{document}


\title{Fine-Tuned Supernova or Failed Explosion? Decoding the Origins of the G3425 Binary}

\correspondingauthor{Zsolt Regály}\email{regaly@konkoly.hu}

\author[0000-0001-5573-8190]{Zsolt Reg\'aly}
\affiliation{HUN-REN Konkoly Observatory, Research Centre for Astronomy and Earth Science, Konkoly-Thege 15-17, 1121, Budapest, Hungary}
\affiliation{CSFK, MTA Centre of Excellence, Budapest, Konkoly Thege Miklós 15-17, H-1121, Budapest, Hungary}

\author[0000-0003-3780-7185]{Viktória Fröhlich}
\affiliation{HUN-REN Konkoly Observatory, Research Centre for Astronomy and Earth Science, Konkoly-Thege 15-17, 1121, Budapest, Hungary}
\affiliation{CSFK, MTA Centre of Excellence, Budapest, Konkoly Thege Miklós 15-17, H-1121, Budapest, Hungary}
\affiliation{ELTE E\"otv\"os Lor\'and University, Institute of Physics, P\'azm\'any P\'eter s\'et\'any 1/A, Budapest, 1117 Hungary}

\author[0000-0001-8764-7832]{József Vinkó}
\affiliation{HUN-REN Konkoly Observatory, Research Centre for Astronomy and Earth Science, Konkoly-Thege 15-17, 1121, Budapest, Hungary}
\affiliation{CSFK, MTA Centre of Excellence, Budapest, Konkoly Thege Miklós 15-17, H-1121, Budapest, Hungary}
\affiliation{ELTE E\"otv\"os Lor\'and University, Institute of Physics, P\'azm\'any P\'eter s\'et\'any 1/A, Budapest, 1117 Hungary}
\affiliation{Department of Experimental Physics, University of Szeged, D\'om t\'er 9, Szeged, 6720, Hungary}

\begin{abstract}
A binary system (G3425) consisting of a massive unseen component and a red giant star on a nearly circular orbit was recently discovered.
The formation of such a system is puzzling because orbital stability generally breaks down due to the large mass loss from the system caused by the SN explosion while forming the unseen component.
Analytical solutions of the variable-mass two-body problem suggest that the explosion should have occurred when the component was close to its apocenter to explain the near-circular remnant system.
This provides a strong constraint on the total mass and orbital configuration of the progenitor system.
The nearly circular orbit of G3425 rules out type II SN scenarios and allows only for a fine-tuned SN~Ib/c explosion to occur when the secondary was close to its apocenter.
Such a scenario, although possible, is highly unlikely.
However, the most likely scenario is a failed SN that produced a black hole, for which no additional constraints on the position of the secondary are needed.
We propose that the unseen component of G3425 is a mass-gap black hole with a mass constrained between the theoretical minimum for failed supernova progenitors ($4~M_\odot$) and the observed upper limit ($4.4~M_\odot$)
Our analysis can be applied to any wide binary system containing an unseen component on a nearly circular orbit.
\end{abstract}


\section{Introduction}
\label{sec:intro}

Recently, \citet{Wang2024NatAs...8.1583W} (W24) discovered an interesting wide binary system, Gaia ID 3425577610762832384 (G3425), by using the Large Aperture Multi-Object Spectroscopic Telescope and data from Gaia DR3.
The mass of the unseen object and the red giant component were derived to be $[2.9-4.4]~M_\odot$ and $[1.7-3.84]~M_\odot$, respectively.
The radius of the red giant turned out to be $[11.2-15.4]~R_\odot$.
The unseen component is more massive than the highest-mass neutron stars (NSs) observed so far. 
Furthermore, its mass is likely to exceed the Tolman--Oppenheimer--Volkoff limit, $[2.2-2.9]~M_\odot$, making it highly probable that this object is a black hole (BH) \citep{Bombaci1996A&A...305..871B,Kalogera1996ApJ...470L..61K}.
Note, however, that in the case of rigidly rotating NSs, the above limit may increase by up to 18-20\%, resulting in a maximum NS mass of 3.48~$M_\odot$ \citep{Cho2018Sci...359..724C,Rezzolla2018ASSL..457.....R}.
The presence of an unseen component means that the binary system experienced a core-collapse supernova explosion. 
G3425 shows an orbital period of $P=877 \pm 2$ days and an eccentricity of $e=0.05\pm0.01$.

The properties of the final remnant of a core-collapse supernova are determined by multiple factors, including progenitor mass, core structure, and explosion energy.
Type II supernovae (SNe~II), which arise from stars that retain their hydrogen envelopes until collapse, are typically associated with red supergiant progenitors.
When the initial mass of the red supergiant is in the range of $[8-20]~M_\odot$, the collapsing iron core is generally small enough that the explosion mechanism (driven by neutrino heating) can succeed, leaving behind a NS \citep{Smartt2009ARA&A..47...63S,Janka2012ARNPS..62..407J, Sukhbold2016ApJ...821...38S}.
For more massive stars, typically above $20~M_\odot$, the likelihood of BH formation becomes significantly higher, either through direct collapse or via the fallback of material onto a newly formed proto--NS that accretes past its stable mass limit \citep{OConnor2011ApJ...730...70O,Ugliano2012ApJ...757...69U, Adams2017MNRAS.469.1445A}.
The typical SN~II progenitors possess extended envelopes with radii of $[500-1500]~R_\odot$ (\citealp{Irani2024ApJ...970...96I}).

Type Ib and Type Ic supernovae (collectively known as stripped-envelope supernovae, SNe~Ib/c) result from progenitors that have lost much or all of their outer hydrogen layers, often via stellar winds or binary interaction.
These events are commonly linked to Wolf–Rayet stars with pre-SN masses of $[4-10]~M_\odot$ \citep{Woosley2006ARA&A..44..507W,Crowther2007ARA&A..45..177C}.
Despite their stripped envelopes, the core-collapse mechanism for SNe~Ib/c is fundamentally similar to SNe~II.
In many cases, the outcome is a NS, but if the stellar core is sufficiently massive, or the explosion energy is marginal, considerable fallback may lead to BH formation \citep{OConnor2011ApJ...730...70O, Sukhbold2016ApJ...821...38S}.
The progenitors of SNe~Ib/c are generally characterized by much smaller radii than SNe~II on the order of $1-10~R_\odot$.

Recent theoretical efforts suggest that failed SNe, where the shock does not revive and minimal or no mass is ejected, generally form BHs.
This process involves either direct collapse or rapid fallback that drives the proto-NS above its stable mass limit  \citep{Zhang2008ApJ...679..639Z,OConnor2011ApJ...730...70O,Ugliano2012ApJ...757...69U,Shariat2025ApJ...983..115S,Sukhbold2016ApJ...821...38S}.
The progenitor stars of these systems are also assumed to be highly stripped He-stars with ZAMS masses of [4-10]~$M_\odot$.
Note that significant observational evidence and theoretical considerations suggest that some BHs experience little to no natal kicks \citep{Mirabel2003Sci...300.1119M,Shenar2022NatAs...6.1085S,Burdge2024Natur.635..316B,Vigna-Gomez2024PhRvL.132s1403V, Nagarajan2025PASP..137c4203N,vanSon2025ApJ...979..209V}
The existence of BHs above $[9-15]~M_\odot$ is also confirmed by both theory and observations \citep{Gerke2015MNRAS.450.3289G,Corral-Santana2016A&A...587A..61C,Adams2017MNRAS.469.1445A,Bahramian2023hxga.book..120B}, however, the minimal birth mass of a BH produced by a failed SN explosion could be just above the NS stability limit.

Studies of the variable-mass two-body problem were initiated by \citet{Hadjidemetriou1963Icar....2..440H,Hadjidemetriou1966Icar....5...34H}, who assumed low mass-loss rates, leading to secular changes only in the semi-major axis, while eccentricity remained constant. 
Later, \citet{Hadjidemetriou1966ZA.....63..116H} demonstrated that extreme mass loss, such as a SN, can drive eccentricity above unity, unbinding the secondary. 
Most subsequent studies assumed adiabatic mass loss with constant eccentricity \citep{debes-sigurdsson2002}. \citet{Veras2011MNRAS.417.2104V} advanced this field by providing analytical and numerical solutions for SN II-like mass-loss regimes using a constant mass-loss rate \citep{Hurley2000MNRAS.315..543H}. 
\citet{Regaly2022ApJ...941..121R} and \citet{Frohlich2023MNRAS.523.4957F} extended these models to binary systems using a simplified homologous SN envelope expansion model.
These studies revealed that secondary components generally acquire high eccentricities and often become unbound.
However, if the mass loss occurs close to the apocenter position of the secondary, the final eccentricity can dampen or even drop to zero.
Since the binary G3425 is almost circular, we can make a reasonable guess about the SN progenitor's properties at the moment of the SN explosion.

In this letter, we present a dynamical study to constrain the orbital properties of the progenitor system of G3425 (i.e., the configuration at the moment of the SN explosion).
We study the change in the orbital elements of the binary systems caused by the mass loss of the primary component, which undergoes either a SN~II, a SN~Ib/c, or a failed SN explosion.

\section{Model calculations}

\begin{figure*}
    \centering
    \includegraphics[width=1\linewidth]{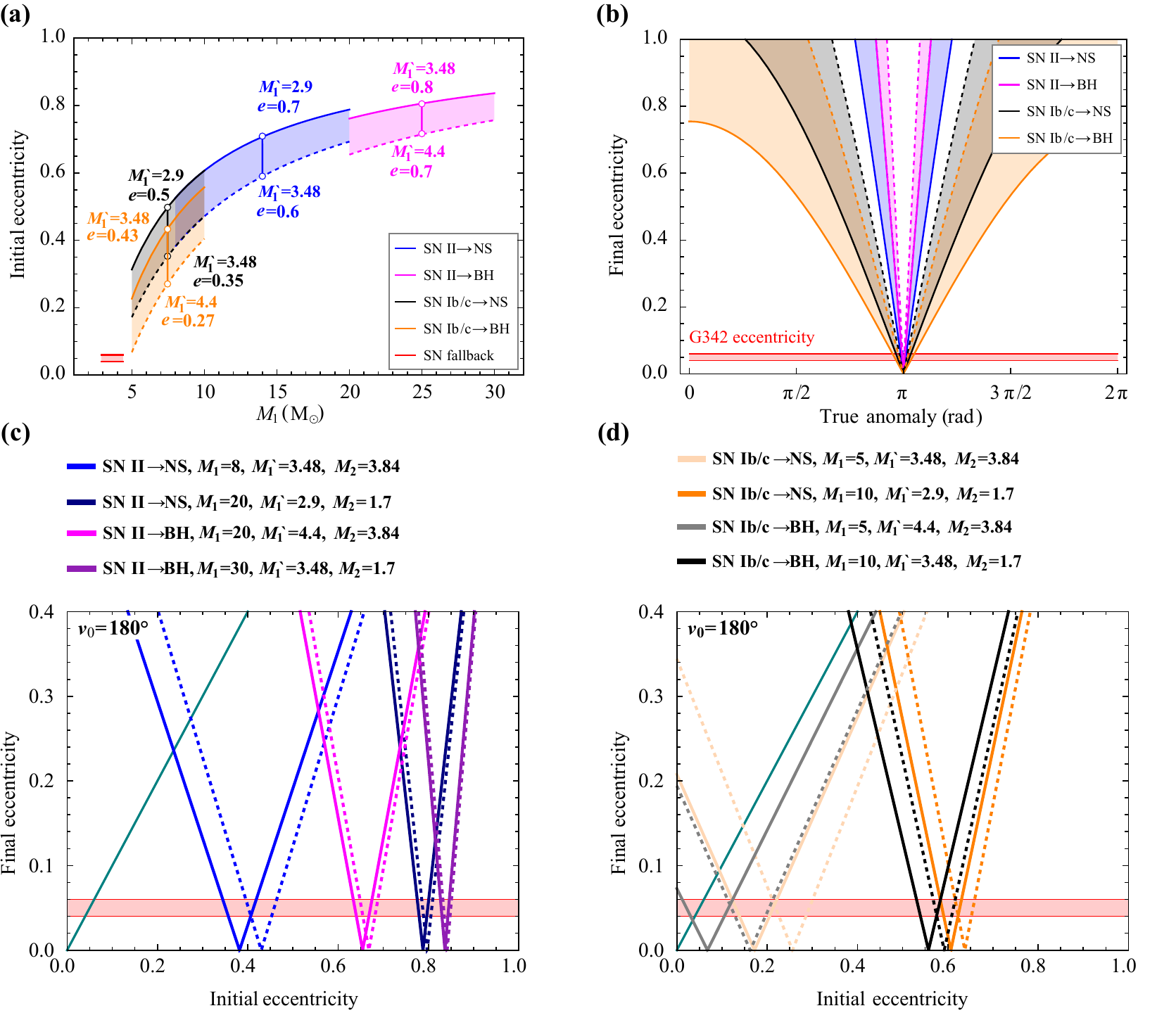}
    \caption{
    Analytical modeling of the G3425 progenitor system.
    In panel (a), the analytical solutions for the initial eccentricity, $e_0$, as a function of the progenitor mass, $M_1$, are shown assuming zero final eccentricity (i.e., $e_1=0$).
    Solid and dashed lines correspond to the minimum and maximum remnant–secondary pairs for a given SN scenario, respectively.
    Plausible progenitor solutions are in the shaded regions for each SN scenario, which are indicated in the legend.
    Panel (b) shows the final eccentricity as a function of the orbital position of the secondary at the moment of the SN explosion using $M_1'$ and $e_0$ values indicated on panel (a). 
    Models with high or low $e_0$ assume $M_2=3.84~M_\odot$ and $M_2=1.7~M_\odot$, respectively.
    The eccentricity of G3425 is shown with a red shaded region.
    Panels (c) and (d) show the final eccentricity as a function of the initial eccentricity of the system for SN~II and SN~Ib/c scenarios, respectively, assuming that the SN explosion occurs when the secondary is at the apocenter.
    The mass parameters for each scenario are shown in the legend.
    Solid lines represent models where the secondary's mass remains constant, while dashed lines represent models where $1~M_\odot$ mass loss is assumed for the secondary due to interaction with the expanding SN envelope.
    The teal line represents the $e_1=e_0$, meaning that eccentricity damping occurs below this line during the SN explosion.}
    \label{fig:e-M1_range}
\end{figure*}

We model five scenarios: 
failed SN explosions with no mass ejection, whose remnant is a BH, SN~II explosions whose remnant is either a NS or a BH, and SN~Ib/Ic explosions whose remnant is a NS or a BH.
For the current masses of the components, we adopt the widest plausible mass ranges from W24: $M_1'=[2.9-4.4]~M_\odot$ for the unseen primary component, and $M_2=[1.7-3.84]~M_\odot$ for the red giant secondary.

In the failed SN scenario, we assume there is no mass loss. Therefore, we assume the progenitor mass is larger than the maximum NS mass, taking into account rotation, and smaller than the observed maximum, $M_1=M_1'=[3.48-4.4]~M_\odot$.
In this case, the orbital elements do not get perturbed: the components inherit the orbital elements of the progenitor system.
As such, no numerical calculation is performed.

In the low-mass SN~II scenario, the progenitor mass is $M_1=[8-20]~M_\odot$, producing a NS with a mass of $M_1=[2.9-3.48]~M_\odot$, assuming a rigidly rotating NS.
In the high-mass SN~II scenario, the progenitor mass is $M_1=[20-30]~M_\odot$, producing a BH of mass $M_1'=[3.48-4.4]~M_\odot$.
In the SN~Ib/c scenario, the progenitor mass is $M_1=[5-10]~M_\odot$, producing either a NS with a mass of $M_1'=[2.9-3.48]~M_\odot$ or a BH of mass $M_1'=[3.48-4.4]~M_\odot$.
Mass transfer or common envelope evolution are not considered, as W24 has firmly ruled out these possibilities: they derive that the companion is far from filling its Roche lobe, its temperature is too low to be a stripped star and there is no additional light source signaling accretion. Binary and spectral synthesis codes also do not find traces of binary interaction, while the binary orbit is too wide for common envelope evolution. Numerical simulations of adiabatic mass loss also necessitate a highly effective ejection from the common envelope.

\subsection{Analytical mass loss model}

In our analytical SN explosion model (see the complete derivation in Appendix~\ref{apx:eccenricity}), instantaneous mass loss is assumed for the SN progenitor.
In this case, the orbital energy of the secondary changes solely due to the instantaneous change in the mass of the primary. 
The semi-major axis and the eccentricity of the system after the explosion are given as
\begin{equation}
    a_1=\frac{\mu_1}{\mu_0}\frac{1}{1+2e_0(\mu_1/\mu_0-1)(1+\cos\nu_0)/(1-e_0^2)},
    \label{eq:sma_fin_maintext}
\end{equation}
\begin{equation}
    e_1 = \sqrt{\frac{\mu_1^2 - \mu_0^2 (1 - e_0^2) - 2 \mu_0 (\mu_1 - \mu_0) (1 + e_0 \cos \nu_0)}{\mu_1^2}},
    \label{eq:e_1_maintext}
\end{equation}
where $\nu_0$ is the true anomaly of the secondary at the onset of the explosion.
$\mu_0$, $a_0$, and $e_0$ are the initial total mass, semi-major axis, and eccentricity, respectively.
The corresponding post-explosion parameters are $\mu_1$, $a_1$, and $e_1$.

Due to mass loss, $a_1$ grows; thus, the progenitor system must have had an orbital separation smaller than the current one before the SN explosion.
However, assuming no binary interaction or common envelope phase, the progenitor orbit could not have been smaller than the size of the SN progenitor star at the onset of the explosion.
This is highly unlikely for SN~II progenitors but guaranteed for SN~Ib/c progenitors, which range in size from $500-1500 R_\odot$ and $1-10 R_\odot$, respectively.

$e_1$ is independent of the binary separation, and can either grow or dampen depending on the orbital position of the secondary at the moment of the SN explosion.
If the secondary is at the apocenter ($\cos \nu_0=-1$), the orbit can be completely circularized (i.e., $e_1=0)$ for a given initial orbital eccentricity.
In this case, the corresponding initial eccentricity can be given as
\begin{equation}
    e_0 = \frac{\mu_0 - \mu_1}{\mu_0}.
    \label{eq:ecc_ini_maintext}
\end{equation}
The extrema of the $e_0$ solutions to the equation $e_1 = 0$ can be defined as
\begin{equation}
    \max\left[e_0\right]=
    \frac{1-\min\left[M_1'\right]/M_1}{\min\left[M_2/M_1\right]-1},
    \label{eq:maxe0_maintext}
\end{equation}    
\begin{equation}  
    \min\left[e_0\right]=
    \frac{1-\max\left[M_1'\right]/M_1}{\max\left[M_2/M_1\right]-1}.
    \label{eq:mine0_maintext}
\end{equation}
The maximum and minimum values of $e_0$ are defined by the minimum and maximum values of the masses $M_1'$ and $M_2$, respectively (panel (a) in Fig.~\ref{fig:e-M1_range}).

It is emphasized that $e_1$ depends on the orbital position of the secondary at the moment of the SN explosion, and only falls in the range observed by W24 when it is close to the apocenter ($\nu_0\simeq180^\circ\pm5^\circ$, see panel (b) in Fig.~\ref{fig:e-M1_range}).
At larger departures from the apocenter, the system gets destabilized (SN~II scenarios) or $e_1$ becomes much larger than what is observed (SN~Ib/c scenarios).
Note that in the failed SN scenarios, there is no change in the binary eccentricity, $e_1=e_0$.

Panels (c) and (d) of Fig.~\ref{fig:e-M1_range} show the final versus initial eccentricity of SN~II and SN~Ib/c models, respectively (mass parameters are shown in the legends).
It is assumed that the SN explosion occurs when the secondary is at its apocenter position.
The solid lines show models where the secondary's mass does not change during the SN explosion.
While increasing the mass of both the remnant and the secondary, the required initial eccentricity for $e_1=0$ also increases:
for the most massive SN~II and SN~Ib/c scenarios, $e_0\gtrsim0.4$ and $e_0\gtrsim0.1$ are required, respectively.
We also investigated models where the secondary has lost $1~M_\odot$ due to interaction with the SN envelope (dashed lines in panels (c) and (d) of Fig.~\ref{fig:e-M1_range}).
As can be seen, the mass loss of the secondary shifts the solutions towards larger initial eccentricities.
Thus, if mass loss occurs, a larger $e_0$ is required to match the observed eccentricity of the binary system.

\subsection{Numerical mass loss model: homologous expansion}
\label{sec:model}

\begin{figure*}
    \centering
    \includegraphics[width=1.0\linewidth]{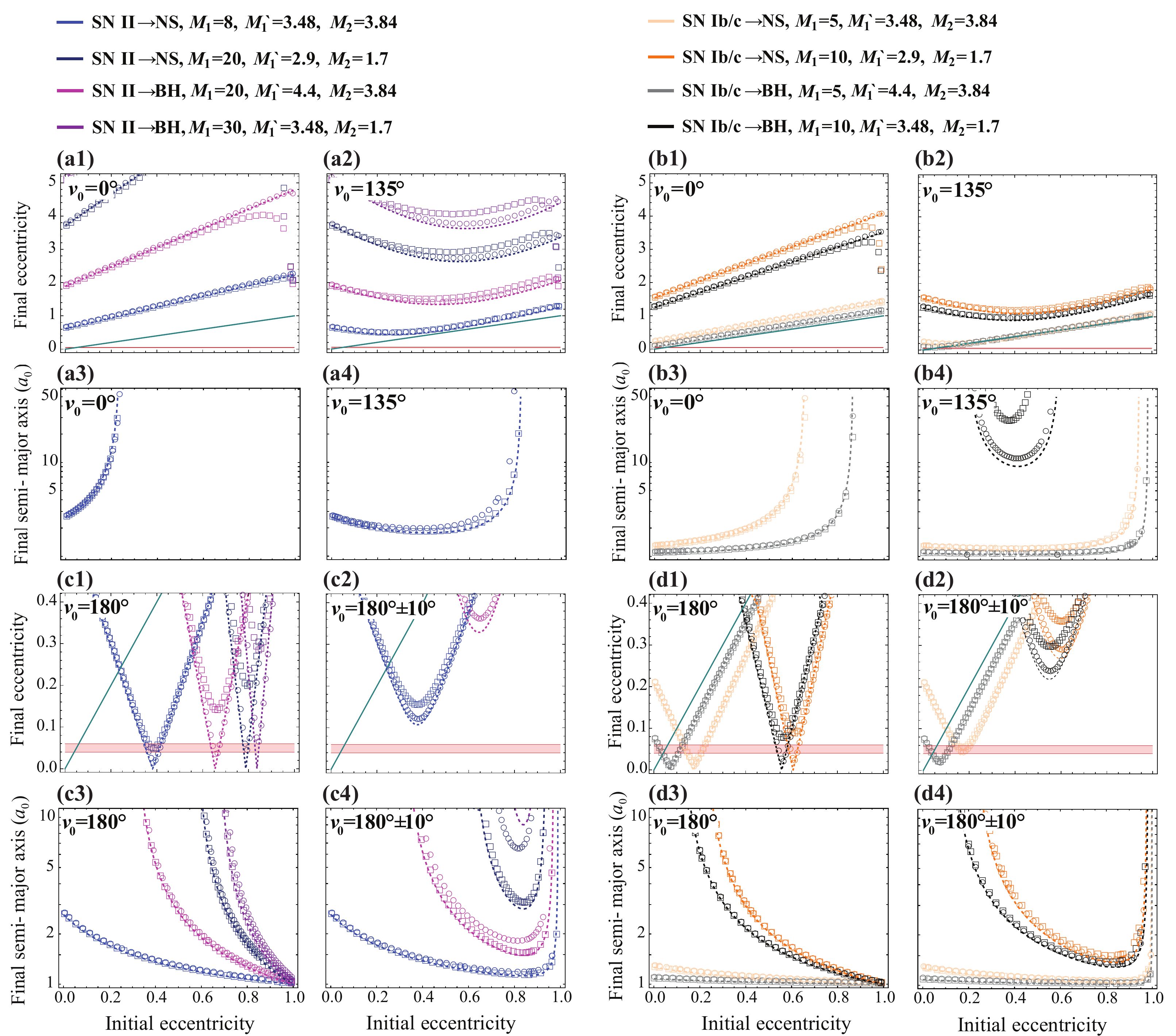}
    \caption{The final versus initial eccentricity and the growth of the semi-major axis of the secondary are illustrated, under the assumption of different unseen remnant masses.
    The left and right plots illustrate SN-II and SN~Ib/c scenarios, respectively.
    Blue and magenta colors represent the heaviest ($M_1'=3.48~M_\odot$, $M_2=3.84~M_\odot$) and lightest ($M_1'=2.9~M_\odot$, $M_2=1.7~M_\odot$) SN~II models, respectively.
    Orange and black colors represent the heaviest ($M_1'=4.4~M_\odot$, $M_2=3.84~M_\odot$) and lightest ($M_1'=3.48~M_\odot$, $M_2=1.7~M_\odot$) SN~II models, respectively.
    Dashed lines correspond to the analytical solution, while symbols represent numerical models.
    Open square and circle symbols represent the homologous envelope expansion model with $6,000~\mathrm{km~s^{-1}}$ and $30,000~\mathrm{km~s^{-1}}$ expansion velocity, respectively.
    Panels illustrate scenarios where the mass loss occurs at pericenter (panels (a1), (a3), (b1), (b3)), intermediate position (panels (a2), (a4), (b2), (b4)), apocenter (panels (c1), (c3), (d1), (d3)), and $\pm10^\circ$ departure from apocenter (panels (c2), (c4), (d2), (d4)).
    The teal-colored line serves as a reference, with dampened and excited eccentricity indicated below and above it, respectively.
    The red shaded regions indicate the observed eccentricity of the secondary $0.05\pm0.01$.
    }
    \label{fig:final-ecc}
\end{figure*}

To elaborate on the instantaneous mass loss assumption, a numerical model is constructed in which the mass inside the secondary's orbit changes due to the envelope loss of the primary by applying a homologous envelope expansion model (see details in Appendix~\ref{apx:homolog}).
While most SN~Ib/c progenitors eject material at velocities below $6,000-15,000~\mathrm{km~s^{-1}}$ \citep{Paragi2010Natur.463..516P,Takaki2013ApJ...772L..17T}, more massive or energetic explosions can result in velocities exceeding $20,000~\mathrm{km~s^{-1}}$ \citep{Maurer2010MNRAS.402..161M,Sanders2012ApJ...756..184S}.
Therefore, we adopt a range of $[6,000-30,000]~\mathrm{km~s^{-1}}$ for the ejecta velocity.

Figure~\ref{fig:final-ecc} shows a comparison of the final eccentricity and semi-major axis of the binary in the analytical and numerical models, assuming four different orbital positions of the secondary in the progenitor system.
The analytical and numerical models match well for all scenarios, assuming a high expansion velocity ($v_\mathrm{max}=30,000~\mathrm{km~s^{-1}}$, circle symbols).
However, at lower ejecta velocity the final eccentricity is larger for scenarios close to the apocenter at a low expansion velocity $v_\mathrm{max}=6,000~\mathrm{km~s^{-1}}$, square symbols), see panels (c1), (c2), (d1), (d2)
As a general phenomenon, the required initial eccentricity for zero final eccentricity shifts toward higher values for the less massive models where the secondary mass, $M_2$, and the remnant mass, $M'_1$, are the smallest, see panels (b1)-(b4) and (d1)-(d4)

Assuming a pericenter position for the secondary at the SN explosion, $e_1$ always grows and the binary system becomes unbound for the majority of models (panels (a1) and (b1)).
In an intermediate position ($\nu_0=135^\circ$) $e_1$ still increases for the majority of models (panels (a2) and (b2)).
At the apocenter position (panels (c1), (d1)), the models can show eccentricity damping.
Thus, there is a certain range of $e_0$ for which $e_1$ matches the observations.
In the vicinity of the apocenter ($\nu_0=180^\circ \pm10^\circ$), there are no SN~II solutions (panel (c2)), and plausible SN~Ib/c solutions emerge only for the low-mass progenitor scenarios (panel (d2)).
In models that match the current observed eccentricity, the growth rate of the binary separation also remains low, $a_1/a_0\lesssim 1.5$ (panels (d3) and (d4)).
This is most evident in the low-mass progenitor and high-mass secondary scenarios, where $M_1=5~M_\odot$ and $M_2=3.84~M_\odot$ (light colors in panels (d3) and (d4)).

The pericenter distance of the secondary derived at the minimum initial eccentricity ($\gtrsim0.4$ for the low-mass and $\gtrsim0.6$ for the high-mass SN~II scenario) is smaller than the SN~II progenitor radius ($\gtrsim500~R_\odot$).
As a result, the secondary would be engulfed by the SN II progenitor.
However, in SN Ib/c scenarios, the progenitor radius ($\lesssim 10 R_\odot$) and the required $e_0$ place the secondary's pericenter far outside the SN progenitor.
Because the growth rate of the binary separation is limited ($a_1/a_0\lesssim 1.5$), the progenitor system was also a wide binary that did not undergo binary interactions, as also hypothesized by W24.
Note that in this case the H and He shells of the SN Ib/c progenitor should have blown off by stellar winds during single-star evolution.
The effectiveness of such winds depends on the metallicity of the star. 
However, the observed solar metallicity of G3425 is consistent with an SN Ib/c event that does not involve binary stripping.

As shown in Fig~\ref{fig:final-ecc}, SN~Ib/c scenarios with a high progenitor mass ($M_1=10~M_\odot$) and a low secondary mass ($M_2=1.7~M_\odot$) require an initial eccentricity of $\simeq0.6$, while scenarios with a low progenitor mass ($M_1=5~M_\odot$) and a high-mass secondary ($M_2=3.84~M_\odot$) match the observations when the initial eccentricity is $\simeq0.1-0.2$.
The minimum final eccentricity that the system can have after the explosion increases with the apocenter distance of the secondary.
Systems with heavier SN progenitors and smaller secondaries experience a greater increase in the minimum final eccentricity.

\subsection{Caveats}

\begin{figure*}[ht!]
    \centering
    \includegraphics[width=1\linewidth]{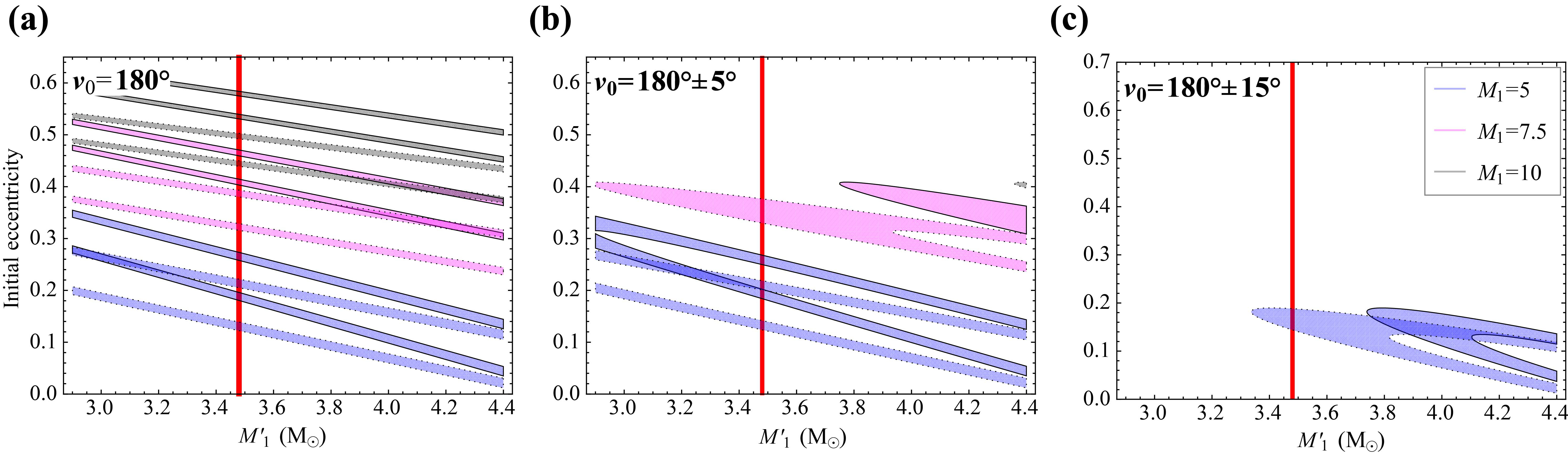}
    \caption{Plausible analytical solutions for the initial eccentricity of the system in SN~Ib/c scenarios as a function of the mass of the unseen component.
    Three different positions of the secondary, $\nu_0=180^\circ,~180^\circ\pm5^\circ,~180^\circ\pm15^\circ$ were assumed.
    The distinct colors represent the three progenitor masses $M_1=5,~7.5,~10~M_\odot$.
    The solid and dotted border regions represent $M_2=1.7$ and $3.83~M_\odot$, respectively.
    The critical mass between NS and BH formation is represented by a red line.
    A movie showing the plausible solutions in a wider range of $\nu_0=180^\circ\pm45^\circ$ is presented in the supplementary material.}
    \label{fig:SNIbc-sol}
\end{figure*}

Here, we mention some caveats of our models.
Although we use an elaborate mass-loss approximation, the homologous expansion model intrinsically assumes spherical symmetry.
Asymmetric envelope ejection can further complicate the perturbation of the orbital elements of the binary \citep{parriott-alcock, namouni, namouni-zhou}.
Local perturbations in the envelope mass distribution by the secondary can break the assumed spherical symmetry.
As a result, both the secondary and primary orbital elements might change.
In a more sophisticated model, the secondary-envelope interactions \citep{debes-sigurdsson2002} could result in mass-loss of the secondary.
As seen in panels (c) and (d) in Fig.~\ref{fig:e-M1_range}, the effect of a $1~M_\odot$ mass loss of the secondary increases the initial eccentricity requirement for a circular final orbit by a few percent.
However, the post-explosion eccentricity damping due to the interaction of the SN ejecta and the secondary was not investigated.
This is a reasonable assumption, as the time of interaction is limited to a short period of time, about $0.1$~yr (Fig.~\ref{fig:Min-t}).

One must also consider the eccentricity damping caused by the tidal forces.
In the constant phase-lag model, the eccentricity-damping timescale is 
\begin{equation}
	\tau_{\mathrm{e}} = 
	\frac{2}{21}
	\frac{Q}{k_2}
	\left ( \frac{a^3}{GM_1’} \right )^{1/2}
	\left ( \frac{M_2}{M_1’} \right )
	\left ( \frac{a}{R_2} \right )^5,
\end{equation}
where $k_2$ is the Love number, $Q$ the tidal quality factor, and $R_2$ the radius of the secondary (\citealp{goldreichsoter1966Icar....5..375G, murrayDermott}).
Assuming a $Q$ range of $10^5-10^{6}$ and a $k_2$ value of $0.1$ \citep{claret2023A&A...674A..67C, smithetal2025A&A...693A.128S}, and taking the widest range of possible masses, $\tau_\mathrm{e}\simeq[26-10,700]$~Gyrs at the current orbital period of 880~days.
Note that even at a smaller separation (corresponding to $a_1/a_0=1.5$ in the SN~Ib/c scenario), this dampening timescale is still significant, $\tau_\mathrm{e}\simeq[1.86-767]$~Gyrs.
The above timescales are in all cases longer than the lifetime of the red giant component, which is $\lesssim1.7$~Gyrs (based on the PARSEC stellar evolution models, \citealp{costaa2019A&A...631A.128C, costab2024eas..conf.2572B, nguyen2022A&A...665A.126N}).

The analysis presented here can be applied to any nearly circular wide binary system containing an unseen object that has been formed in a core collapse SN explosion. 
X-ray binaries with mass gap BHs 4U~1543-47 \citep{Orosz1998ApJ...499..375O}, GX~339-4 \citep{Heida2017ApJ...846..132H}, and J05215658 \citep{Thompson2019Sci...366..637T} all have nearly circular orbits.
Note, however, that in X-ray binaries, the secondary is so close to the compact object that they are usually strongly interacting.
In addition, since the overflow of the Roche lobe occurs before the explosion of SN~Ib/c \citep{taurisetal17,yoonetal10}. 
Thus, it is difficult to derive the parameters for the most plausible progenitor system of close X-ray binaries.

In triple star systems, the canonical formation channel for circular low-mass X-ray binaries involves the inner binary forming a BH+RG system, in which Kozai--Lidov resonances and tides can effectively shrink and circularise the orbit \citep{Naoz2016ApJ...822L..24N, Shariat2025ApJ...983..115S}.
If the triple system loses its massive component during triple common envelope evolution, the probability of the eccentricity growing from zero to the observed value of 0.05 during an SN~Ib/c explosion is $\simeq50$\% \citep{Li2025ApJ...979L..37L}.
According to our results, a zero-eccentricity binary system that undergoes an SN~Ib/c explosion cannot reproduce the orbit of G3425 unless its pre-explosion eccentricity is $\gtrsim0.1$ and the explosion happens near apocenter.
This discrepancy needs to be further investigated in the future.

\section{Summary and Conclusion}
\label{sec:conclusion}

Plausible progenitor systems accounting for the observed eccentricity and semi-major axis of the G3425 remnant system can be constructed based on our models.
First, in the failed SN scenario, we assume that there is no mass loss, thus, the orbital parameters of the secondary do not change (unless the envelope surpasses the orbit before the fallback).
Therefore, this scenario is readily able to reproduce the observed very low eccentricity of G3425.
The only constraint for the progenitor system is that the progenitor mass was relatively small $[3.48-4.4]~M_\odot$ and the eccentricity of the secondary was $0.05\pm0.01$.

However, none of the SN~II scenarios can reproduce the eccentricity of G3425 at the necessary semi-major axis.
This is because the required initial eccentricity of the binary system ($\gtrsim 0.4$) is so high that the progenitor would engulf the secondary (see Fig.~\ref{fig:final-ecc}).
The assumption that the secondary must orbit outside the envelope is made here, since surviving inside the stellar envelopes will likely commence unstable mass transfer, and thus survival chances are low \citep{bear-etal}.

Nevertheless, SN~Ib/c scenarios can also provide plausible progenitor system configurations.
Analytical solutions for three different progenitor masses ($M_1=5,~7.5,~10~M_\odot$) are shown in Fig.~\ref{fig:SNIbc-sol}.
Solutions can match the eccentricity of the observed system if the progenitor eccentricity is in the range of $0.01\lesssim e_0 \lesssim 0.65$.
However, only limited regions of the $M_1'-e_0$ plane give valid solutions.
As the secondary departs from the apocenter, the range of plausible eccentricities shrinks.
There is no solution for $M_1=10~M_\odot$ if the secondary was away from its apocenter by $\pm5^\circ$ at the onset of SN explosion (panel (b)).
Further departing from the apocenter by $\pm15^\circ$ means that the models with $M_1\geq7.5M_\odot$ are unable to explain the observations, and the plausible remnant mass is above the NS limit.
The probability that the secondary is at $\nu_0 = 180^\circ$, $\pm 5^\circ$, $\pm 10^\circ$, or $\pm 15^\circ$ is equally $\lesssim10\%$.
This is because the probability that the secondary is at a given $\nu_0=180^\circ \pm \Delta \nu$ range is proportional to the initial eccentricity and $\Delta \nu$ (see details in Appendix~\ref{apx:prob}), while the maximum value of the initial eccentricity that gives a matching solution to the observations decreases with $\Delta \nu$ (Fig.~\ref{fig:SNIbc-sol}).
Consequently, the overall probability remains almost the same for different values of $\Delta \nu$.

Assuming that the progenitor system of G3425 was a stripped helium-rich or completely stripped helium-free star with a mass of $[4-10]~M_\odot$ that exploded as a SN Ib/c, the parameters of G3425 can be reproduced in our models.
In this case, the SN remnant may be either a NS or a BH.
The originally eccentric orbit of the binary progenitor matches the observed low eccentricity of the remnant system G3425 only if the explosion occurred close to the apocenter, which has a probability of $\lesssim10\%$.
A SN~Ib/c explosion occurring at a larger departure from the apocenter position favors higher remnant mass, i.e., the unseen component is likely a BH.
Furthermore, the homologous expansion models showed that plausible solutions for the progenitor system favor a large envelope expansion velocity $\gtrsim10,000~\mathrm{km~s^{-1}}$.

A failed SN explosion where no mass loss occurred can also explain the nearly circular G3425 system if the progenitor eccentricity was in the range of the observed one.
In this case, no fine-tuning of the eccentricity or true anomaly of the model is required, and the remnant is most likely a BH, since the progenitor mass must have been greater than $4~M_\odot$.
Based on the above considerations, it is highly probable that the G3425 system indeed hides a mass-gap BH of $[4-4.4]~M_\odot$.

\begin{acknowledgments}
This research is supported by the projects NKFIH OTKA K142534 and GINOP 2.3.2-15-2016-00033.
VF is supported by the undergraduate research assistant program of the Konkoly Observatory.
The authors acknowledge Sz. Csizmadia and A. Smith for the fruitful discussion on stellar Love numbers and tidal parameters.
We thank the anonymous referee for significantly improving the quality of the paper.
\end{acknowledgments}

\clearpage
\appendix

\section{Eccentricity excitation and orbital expansion in an instantaneous mass loss model}
\label{apx:eccenricity}

When a star in a binary system loses mass instantaneously, the orbit of the secondary changes, altering both the semi-major axis and the orbital eccentricity.
The eccentricity after the mass loss is influenced by the specific orbital energy and angular momentum conservation.
The following derivation considers that the mass loss is rapid enough to treat the secondary’s position and velocity as constants during the event.
Before the explosion, the secondary orbits the primary in an elliptical orbit characterized by an initial semi-major axis $a_0$ and eccentricity $e_0$. The true anomaly at the moment of mass loss is denoted as $\nu_0$.

The binary star system consists of a primary star with mass $M_1$ and a secondary star with mass $M_2$, where $M_1 > M_2$. 
The total initial mass of the system is given by
\begin{equation}
    \mu_0 = M_1 + M_2.
\end{equation}
At a given time, the primary star instantaneously loses mass such that its new mass is $M_1'$. The new total mass of the system is
\begin{equation}
    \mu_1 = M_1' + M_2.
\end{equation}
Since the mass loss is assumed to be instantaneous, the secondary's position and velocity remain unchanged at the moment of mass loss. 
Moreover, here it is assumed that the secondary's mass is not changed during the SN explosion, namely $M_2$ is constant.
We aim to determine the new semi-major axis $a_1$ and eccentricity $e_1$ of the system in terms of $\mu_0$, $\mu_1$, $e_0$, and $\nu_0$.

The radial distance of the secondary from the primary, $r$, at true anomaly $\nu_0$ is given by the equation of an elliptical orbit
\begin{equation}
    r = \frac{a_0 (1 - e_0^2)}{1 + e_0 \cos \nu_0}.
    \label{eq:ellipticorbit}
\end{equation}
The velocity of the secondary is determined from the conservation of energy,
\begin{equation}
    v^2 = \mu_0 \left(\frac{2}{r} - \frac{1}{a_0} \right).
    \label{eq:vis-visa1}
\end{equation}
After the mass loss, energy conservation gives the 
\begin{equation}
    v^2 = \mu_1 \left(\frac{2}{r} - \frac{1}{a_1} \right).
    \label{eq:vis-visa2}
\end{equation}
Thus, using Eqs.~(\ref{eq:vis-visa1})-(\ref{eq:vis-visa2}) and the equation of the elliptical orbit, Eq.~(\ref{eq:ellipticorbit}), the new semi-major axis, $a_1$, can be given as
\begin{equation}
    a_1=\frac{\mu_1}{\mu_0}\frac{1}{1+2e_0(\mu_1/\mu_0-1)(1+\cos\nu_0)/(1-e_0^2)}.
    \label{eq:sma_fin}
\end{equation}

\begin{figure*}
    \centering
    \includegraphics[width=1\linewidth]{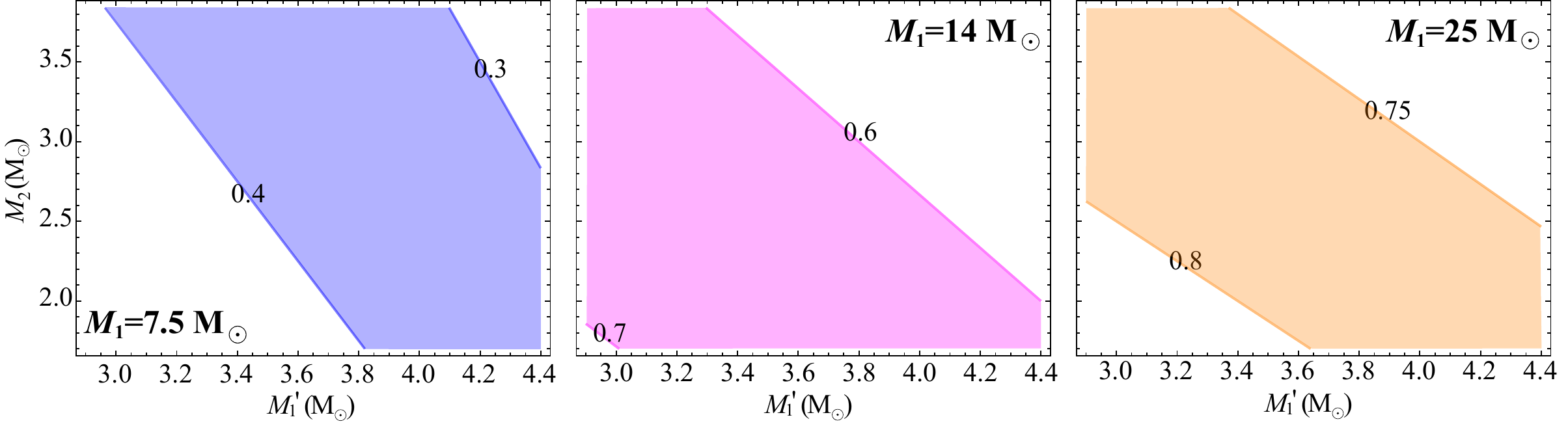}
    \caption{Ranges of initial eccentricity, $e_0$, calculated according to Eq.~(\ref{eq:ecc_ini}) on the $M_1'-M_2$ plane assuming three different progenitor masses of $M_1=7.5,~14$ and $25~M_\odot$.
    Plausible solutions for $e_0$ at a given $M_1'$, $M_2$ are in the shaded regions.}
    \label{fig:M1'-M2}
\end{figure*}

To derive the new eccentricity, we consider the specific angular momentum before and after the mass loss ($h_0$ and $h_1$, respectively):
\begin{equation}
    h_0^2 = \mu_0 a_0 (1 - e_0^2),~~~h_1^2 = \mu_1 a_1 (1 - e_1^2).
\end{equation}
Since angular momentum is conserved, we equate these expressions and solve them for $e_1$, which gives
\begin{equation}
    e_1^2 = 1 - \frac{\mu_0 a_0 (1 - e_0^2)}{\mu_1 a_1}.
\end{equation}
Using Eqs.~(\ref{eq:sma_fin}) and (\ref{eq:ellipticorbit}) this can be expressed as
\begin{equation}
    e_1 = \sqrt{\frac{\mu_1^2 - \mu_0^2 (1 - e_0^2) - 2 \mu_0 (\mu_1 - \mu_0) (1 + e_0 \cos \nu_0)}{\mu_1^2}}.
    \label{eq:ecc_fin}
\end{equation}
This final expression provides the new eccentricity as a function of the initial and new mass, initial eccentricity, and initial true anomaly of the system.
The true anomaly plays a critical role because the secondary’s position determines whether the change in the gravitational force enhances or reduces the secondary's radial velocity, directly affecting the orbital shape.
If the mass loss occurs near pericenter $(\nu_0 = 0^\circ$), the secondary experiences a sharp reduction in gravitational pull at a point where its velocity is highest, leading to a more elliptical orbit.
Conversely, if the mass loss happens near apocenter $(\nu_0 = 180^\circ$), where the secondary's velocity is lowest, the orbit may become less eccentric or even circularize in some cases.
In this case, the eccentricity equation, Eq.~(\ref{eq:ecc_fin}), is employed, and the parameter $e_1$ is set to zero, resulting in the following equation
\begin{equation}
    \frac{\mu_0(e_0-1) + \mu_1}{\mu_1}= 0.
\end{equation}
Rearranging this equation yields:
\begin{equation}
    e_0 = \frac{\mu_0 - \mu_1}{\mu_0}.
    \label{eq:ecc_ini}
\end{equation}
This solution indicates that the initial eccentricity depends on the fractional mass loss $(\mu_0 - \mu_1)$, the initial total mass $\mu_0$.
The extrema of $e_0$ as defined by Eq.~(\ref{eq:ecc_ini}) can be given as 
\begin{equation}
    \max\left[e_0\right]=\frac{\max\left[1-M_1'\right]/M_1}{\min\left[M_2/M_1-1\right]}=\frac{1-\min\left[M_1'\right]/M_1}{\min\left[M_2/M_1\right]-1},~~~\min\left[e_0\right]=\frac{\min\left[1-M_1'\right]/M_1}{\max\left[M_2/M_1-1\right]}=\frac{1-\max\left[M_1'\right]/M_1}{\max\left[M_2/M_1\right]-1}.
    \label{eq:extrema}
\end{equation}
Thus, the widest plausible initial eccentricity range belongs to mass pairs of $\min[M_1',M_2]$ and $\max[M_1',M_2]$.
Figure~\ref{fig:M1'-M2} shows ranges of plausible $M_1'$ and $M_2$ values for $M_1=7.5, 14$ and $25~M_\odot$ (the mean mass values for the progenitor for each scenario investigated), for which case the eccentricity of the remnant system is zero.
Contour lines show the $e_0$ values for the $e_1=0$ solution at the extrema.

\section{Homologous expansion model}
\label{apx:homolog}

In the following, we summarize the homologous expansion model of the stellar envelope \citep[e.g.][]{arnett80, bw17, vinko04, Regaly2022ApJ...941..121R, Frohlich2023MNRAS.523.4957F}
Homologous expansion means that 1) the velocity of the ejected layers is a linear function of the distance from the star; 2) the velocity of the outermost layer, $v_{\mathrm{max}}$, is constant, and; 3) the density profile of the ejecta is also time-independent.
We assume that the inner 10\% of the star contains a constant density core, which has a radius $r_\mathrm{c}=0.1R$.
Initially, the radius of the envelope ($R$) coincides with the radius of the progenitor star.
We assume a progenitor size of $500~R_\odot$ and $10~R_\odot$ for the SN~II and SN~Ib/c cases, respectively.

The expansion velocity of the envelope at a distance $r$ is $v(r,t) = \left(r(t)/R(t) \right) v_\mathrm{max}$, where $R(t)$ is the radius of the outermost layer of the ejecta and $v_\mathrm{max}$ is its expansion velocity.
We introduce the co-moving distance as $x=r(t)/R(t)$, so envelope velocity at $x$ is $v(x,t)=x \cdot v_\mathrm{max}$.
Defining $\Delta t$ as the elapsed time, the distance of a mass shell from the SN center is
\begin{equation}
    r(t) = r(0) + v(r,t) \Delta t = x(R_\mathrm{0}+v_\mathrm{max} \Delta t).
    \label{eq:envelope_r}
\end{equation}
When this shell reaches the position of the secondary, $r = r_\mathrm{sec}$, thus 
\begin{equation}
    x(r_\mathrm{sec}) \equiv x_\mathrm{sec}=\frac{r_\mathrm{sec}}{R_\mathrm{0}+v_\mathrm{max} \Delta t}.
    \label{eq:x_p}
\end{equation}

When determining the mass inside the orbit of the secondary, $M_\mathrm{in}$, three cases are distinguished:
1) $x_\mathrm{sec}>1$; 2) $x_\mathrm{sec} > x_\mathrm{c}$ and; 3) $x_\mathrm{sec}<x_\mathrm{c}$
In the first case, all of the progenitor's mass resides within the secondary's orbit, so $M_\mathrm{in}=M_\mathrm{ej}+M_\mathrm{r}$, where $M_\mathrm{ej}$ is the mass of the ejecta, and $M_\mathrm{r}$ the mass of the remnant.
In the second and third cases, we need to take into account the change in envelope density.
The density of the core, $\rho_0$, is assumed to be spatially constant, while the density of the envelope follows a power-law,
\begin{equation}
    \rho=\rho_0 \left(\frac{x}{x_\mathrm{c}}\right)^{-n},
    \label{eq:density}
\end{equation}
where $n=7$ is assumed when $x > x_\mathrm{c}$, and $n=0$ otherwise.
The core density changes in time according to
\begin{equation}
    \rho_0(t) = \frac{3 M_\mathrm{c}}{4 \pi} \left ( x_\mathrm{c} (R_\mathrm{0}+v_\mathrm{max} \Delta t) \right )^{-3}.
    \label{eq:coredens}
\end{equation}
If $r_\mathrm{sec} > r_\mathrm{c}$, the mass residing within the secondary's orbit is
\begin{equation}
    M_\mathrm{in}= M_\mathrm{n} + M_\mathrm{c} + \int_{r_\mathrm{c}}^{r_\mathrm{sec}} 4 \pi r^2 \rho(r) dr,
    \label{eq:min_def}
\end{equation}
where $r_c$ denotes the radius of the core.
If $r_\mathrm{sec}>r_\mathrm{c}$, the above equation gives
\begin{equation}
    M_\mathrm{in}=M_\mathrm{n} + M_\mathrm{c}+ 4 \pi \rho_\mathrm{0} R_\mathrm{0}^3 x_\mathrm{c}^n \int_{x_\mathrm{c}}^{x_\mathrm{sec}} x^{2-n} dx
    = M_\mathrm{n} + M_\mathrm{c} \left [ 1+ \frac{3}{n-3} \left ( 1- \left ( \frac{x_\mathrm{sec}}{x_\mathrm{c}} \right )^{3-n} \right ) \right ].    
    \label{eq:min_xp>xc}
\end{equation}
On the other hand, if $x_\mathrm{sec}<x_\mathrm{c}$,
\begin{equation}
    M_\mathrm{in}=M_\mathrm{n}+ 4 \pi \rho_\mathrm{0} R_\mathrm{0}^3  \int_{0}^{x_\mathrm{sec}} x^2 dx =
    M_\mathrm{n} + M_\mathrm{c} \left ( \frac{x_\mathrm{sec}}{x_\mathrm{c}} \right )^3.
    \label{eq:min_xp<xc}
\end{equation}
Using the aforementioned density profile, ejecta mass can be expressed as
\begin{equation}
    M_\mathrm{ej} = 4 \pi R_0^3 \rho_\mathrm{0} \left ( \int_{0}^{x_\mathrm{c}} x^2 dx + x_\mathrm{c}^n \int_{x_\mathrm{c}}^{1} x^{2-n} dx \right ) = 
    4 \pi R_0^3 \rho_\mathrm{0} \left ( \frac{x_\mathrm{c}^3}{3} + \frac{x_\mathrm{c}^n-x_\mathrm{c}^3}{3-n} \right ).
    \label{eq:mej}
\end{equation}
From the above, we derive the core density as
\begin{equation}
    \rho_\mathrm{0} = \frac{M_\mathrm{ej}}{4 \pi R_0^3} \left ( \frac{x_\mathrm{c}^3}{3} + \frac{x_\mathrm{c}^n-x_\mathrm{c}^3}{3-n} \right )^{-1}.
    \label{eq:rho0}
\end{equation}
Thus, core mass can be calculated as
\begin{equation}
    M_\mathrm{c}=\frac{4 \pi R_0^3 x_\mathrm{c}^3}{3} \rho_\mathrm{0}= \frac{M_\mathrm{ej}} {1+ \frac{3}{n-3} \left( 1- x_\mathrm{c}^{n-3} \right ) }.
    \label{eq:mcore}
\end{equation}

\begin{figure*}
    \centering
    \includegraphics[width=1\linewidth]{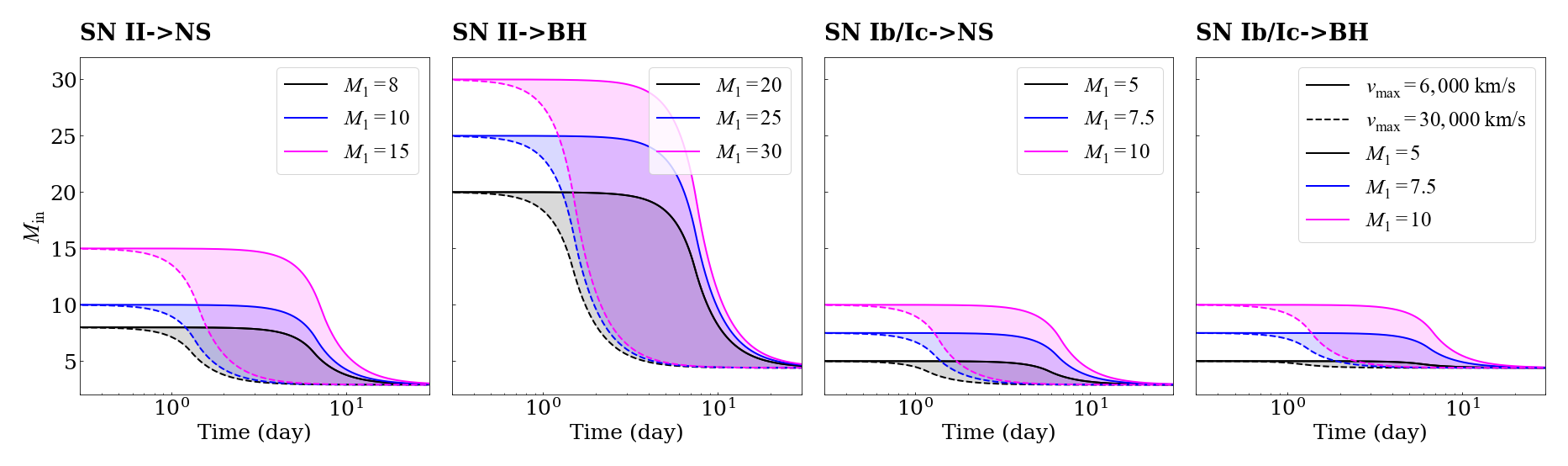}
    \caption{The mass inside the secondary's orbit, $M_\mathrm{in}$, as a function of time in the four core collapse SN scenarios.
    Three different progenitor masses are assumed in each scenario, as indicated by the colors in the legend.
    Solid and dashed lines correspond to expansion velocities of $6,0000$ and $30,0000~\mathrm{km~s^{-1}}$, respectively.}
    \label{fig:Min-t}
\end{figure*}

Figure~\ref{fig:Min-t} shows $M_\mathrm{in}$ as a function of time for four SN scenarios.
As can be seen, the mass inside the secondary's orbit decreases rapidly.  
In about a month, there will be no significant mass between the secondary and the remnant.
Therefore, if there is any interaction between the expanding envelope and the secondary, it is only temporary.

To model the perturbations in the orbital elements of the binary, we solve the equations of motion numerically in two dimensions.
The mass inside the secondary's orbit, $M_\mathrm{in}$, changes due to the envelope loss of the primary; thus, $M_\mathrm{in}$ needs to be updated at every time step according to the homologous expansion.
Numerical integration is done in Python 3.8.8 using SciPy's \verb|integrate.solve_ivp()| function with an explicit 8th-order Runge-Kutta method with an adaptive step size.
Integration time for each system is five million days (about 13,700 years), which allows for the monitored orbital elements to relax by the end of the simulation.

\section{Probability of being at apocentre} 
\label{apx:prob}

\begin{figure}
    \centering
    \includegraphics[width=0.8\linewidth]{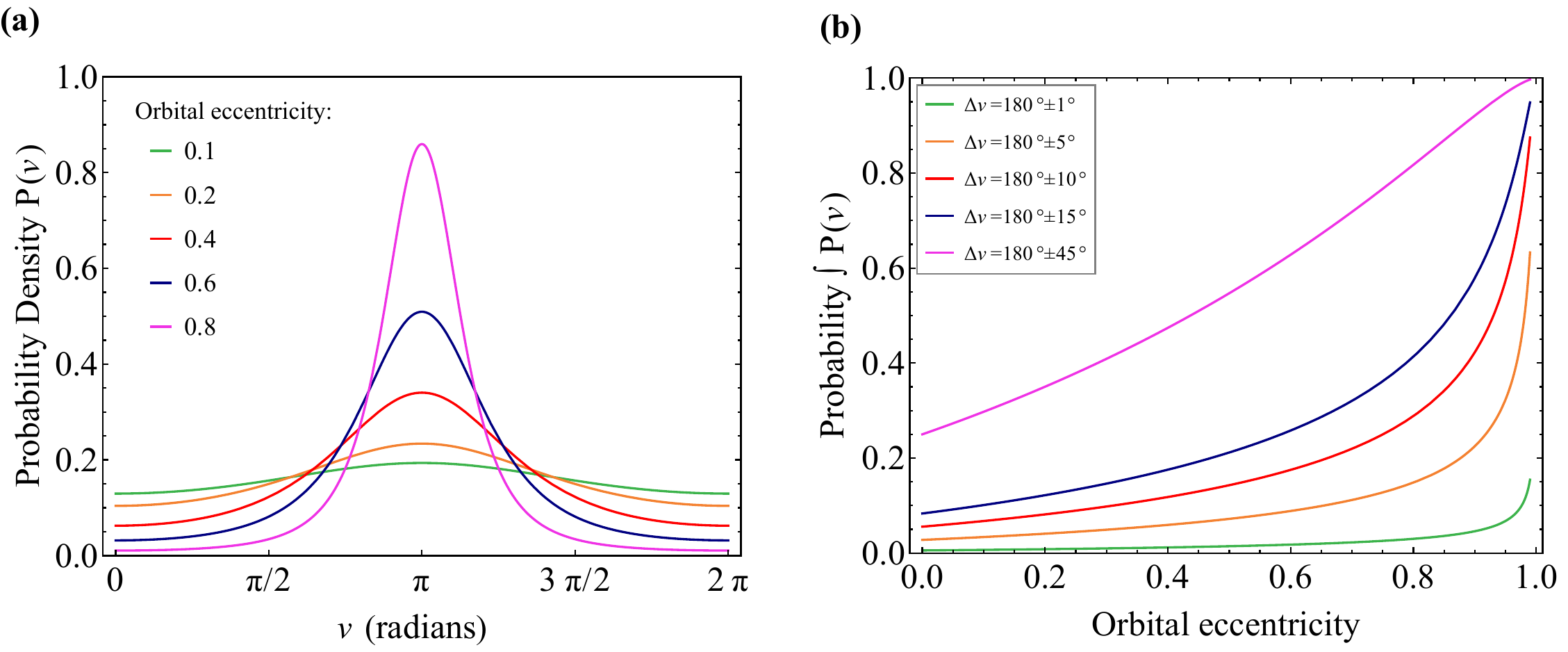}
    \caption{Panel (a): Probability density functions given by Eq.~(\ref{eq:prob}) for various eccentricities shown with different colors.
    Panel (b): Probability that the secondary is found at a given true anomaly range, as a function of orbital eccentricity.}
    \label{fig:prob}
\end{figure}

In classical two-body orbital mechanics, the probability of finding a small body at a given true anomaly $\nu$ over one orbital period can be derived by noting that the mean anomaly $M$ increases uniformly in time, while the transformation from $M$ to $\nu$ depends on the eccentricity $e$.
Following the standard treatments of \citet{murrayDermott}, one starts with Kepler’s second law, which implies a uniformly swept area and thus a uniform distribution of $M$ in time.
Subsequent application of the identities linking the mean anomaly $M$, the eccentric anomaly $E$, and the true anomaly $\nu$ yields the probability density
\begin{equation}
P(\nu) \;=\; \frac{1}{2\pi}\,\frac{\bigl(1 - e^2\bigr)^{3/2}}{\bigl(1 + e \,\cos \nu\bigr)^{2}}.
\label{eq:prob}
\end{equation}
Panel (a) of Fig.~\ref{fig:prob} shows the probability density function for various orbital eccentricities.
It is appreciable that the secondary spends more time near apocentre, where its orbital speed is minimal.
The probability that the secondary resides in the range of $[\nu_0-\nu_1]$ can be given as
\begin{equation}
    \int_{\nu_0}^{\nu_1} P(\nu)d\nu=\left[\left(1-e^2\right)^{3/2} \left(\frac{e \sin (\nu )}{\left(e^2-1\right) (e \cos (\nu )+1)}-\frac{2
   \tanh ^{-1}\left((e-1) \tan \left(\frac{\nu
   }{2}\right)/(\sqrt{e^2-1})\right)}{\left(e^2-1\right)^{3/2}}\right)\right]_{\nu_0}^{\nu_1},
   \label{eq:probint}
\end{equation}
as shown on panel (b) of Fig.~\ref{fig:prob} for different $\Delta \nu$ values as a function of the orbital eccentricity.
Table~\ref{tab:prob} displays the probabilities given by Eq.~(\ref{eq:probint}) for three different $\Delta \nu$ values.
The initial eccentricity values are defined by the maximum possible values that can give solutions that match the observations for a given $\Delta \nu$ (Fig.~\ref{fig:SNIbc-sol}).
The probabilities of valid solutions for the G3425 system are shown in boldface.
We emphasize that the calculated probabilities are upper estimates and solutions also exist with smaller initial eccentricities (see Fig.~\ref{fig:SNIbc-sol}), for which case the probabilities are smaller than what is presented in Table~\ref{tab:prob}.

\begin{table}[]
\begin{center}
\caption{Probabilities of finding the secondary at a given $180^\circ \pm \Delta \nu$ range in an orbit with an initial eccentricity of $e_0$.}
\begin{tabular}{cccc}
\hline
\hline
    $e_0$ & $\Delta \nu = 5^\circ$  &$\Delta \nu = 10^\circ$ &$\Delta \nu = 15^\circ$ \\
    0.65 & \textbf{10\%}  & 20\%  & 29\%   \\
    0.4  & \textbf{6\%}   & \textbf{12\%}  & 18\%   \\
    0.2  & \textbf{4\%}   & \textbf{8\%}   & \textbf{12\%}   \\
    \hline
\end{tabular}
\label{tab:prob}
\end{center}
\end{table}

\bibliographystyle{aasjournal}
\bibliography{SN-BH}

\end{document}